\documentclass[aps,prl,superscriptaddress,twocolumn,floatfix,longbibliography]{revtex4-1}
\usepackage{dcolumn}
\usepackage{graphicx,color}
\usepackage[utf8]{inputenc}
\usepackage{amsmath,amssymb,bm,amsfonts,dsfont,mathrsfs,amsthm}
\usepackage{braket}
\usepackage{txfonts,comment}
\usepackage[colorlinks=true,linkcolor=blue,citecolor=blue,urlcolor=blue]{hyperref}
\usepackage{soul}

\newcommand{\red}[1]{\textcolor{black}{#1}}
\newcommand{\vm}[1]{\textcolor{black}{#1}}
\newcommand{\blue}[1]{\textcolor{black}{#1}}
\newcommand{\mga}[1]{\textcolor{black}{#1}}

\begin{document}

\title{Enhanced quantum frequency estimation by nonlinear scrambling}

\author{Victor Montenegro}
\email{victor.montenegro@ku.ac.ae}
\affiliation{College of Computing and Mathematical Sciences, Department of Applied Mathematics and Sciences, Khalifa University of Science and Technology, 127788 Abu Dhabi, United Arab Emirates}
\affiliation{Institute of Fundamental and Frontier Sciences, University of Electronic Science and Technology of China, Chengdu 611731, China}
\affiliation{Key Laboratory of Quantum Physics and Photonic Quantum Information, Ministry of Education, University of Electronic Science and
Technology of China, Chengdu 611731, China}

\author{Sara Dornetti}
\email{sara.dornetti@studenti.unimi.it}
\affiliation{Quantum Technology Lab $\&$ Applied Quantum Mechanics Group, Dipartimento di Fisica ``Aldo Pontremoli'', Universit\`a degli Studi di Milano, I-20133 Milano, Italia}

\author{Alessandro Ferraro}
\email{alessandro.ferraro@unimi.it}
\affiliation{Quantum Technology Lab $\&$ Applied Quantum Mechanics Group, Dipartimento di Fisica ``Aldo Pontremoli'', Universit\`a degli Studi di Milano, I-20133 Milano, Italia}

\author{Matteo G. A. Paris}
\email{matteo.paris@fisica.unimi.it}
\affiliation{Quantum Technology Lab $\&$ Applied Quantum Mechanics Group, Dipartimento di Fisica ``Aldo Pontremoli'', Universit\`a degli Studi di Milano, I-20133 Milano, Italia}

\date{\today}

\begin{abstract}
Frequency estimation, a cornerstone of basic and applied sciences, has been significantly enhanced by quantum sensing strategies. Despite breakthroughs in quantum-enhanced frequency estimation, key challenges remain: static probes limit flexibility, and the interplay between resource efficiency, sensing precision, and potential enhancements from nonlinear probes remains not fully understood. In this work, we show that dynamically encoding an unknown frequency in a nonlinear quantum electromagnetic field can significantly improve frequency estimation. To provide a fair comparison of resources, we define the energy cost as the figure of merit for our sensing strategy. We further show that specific higher-order nonlinear processes lead to nonlinear-enhanced frequency estimation. This enhancement results from quantum scrambling, where local quantum information spreads across a larger portion of the Hilbert space. We quantify this effect using the Wigner-Yanase skew information, which measures the degree of noncommutativity in the Hamiltonian structure. Our work sheds light on the connection between Wigner-Yanase skew information and quantum sensing, providing a direct method to optimize nonlinear quantum probes.
\end{abstract}

\maketitle

\textit{Introduction---} Accurate frequency estimation plays a central role in various fields such as quantum communication~\cite{gisin2007quantum, chen2021review}, quantum sensing~\cite{degen2017quantum, pirandola2018advances, montenegro2024reviewquantummetrologysensing}, and quantum computation~\cite{aharonov1999quantum}. Its importance extends to key applications in metrology~\cite{giovannetti2004quantum, giovannetti2006quantum, giovannetti2011advances, udem2002optical, descamps2023quantum, kuenstner2024quantummetrologylowfrequency, boss2017quantum, cai2021quantum}, spectroscopy~\cite{roos2006designer, schmitt2021optimal, lamperti2020optical}, and precision timekeeping~\cite{ludlow2015optical}. Quantum sensors, probes that exploit quantum phenomena, have been demonstrated to surpass the precision limits achievable by classical sensors in frequency estimation~\cite{haase2018fundamental, bollinger1996optimal, degen2017quantum, montenegro2024reviewquantummetrologysensing,giovannetti2004quantum, giovannetti2006quantum, giovannetti2011advances}. 
\mga{Thus, the task of quantum frequency estimation has been pursued in various scenarios, including the use of decoherence-free subspaces for trapped particles in optical lattices~\cite{dorner2012quantum}, continuous monitoring of qubit probes in noisy environments~\cite{albarelli2018restoringheisenberg,albarelli2020quantum}, techniques involving adaptive coherent control~\cite{naghiloo2017achieving, rodriguez2024usefulness, gutierrezrubio2021optimalfrequencyestimationapplication}, and general scenarios in noisy metrology~\cite{huelga1997improvement, smirne2016ultimate, haase2016precision, macieszczak2014bayesian, macieszczak2015zeno,riberi2022frequency}. Although quantum-enhanced sensing strategies have demonstrated remarkable precision in frequency estimation~\cite{donohue2018quantum, frowis2014optimal}, several key questions remain. In particular, static methods based on the preparation of a given state do not allow for effective encoding, as the dependence on frequency of the eigenvalues and eigenvectors leads to unfavourable scaling. This limits the achievable precision. We therefore ask: \blue{(i) Is it possible to remove this restriction using dynamical approaches, thereby increasing the flexibility of the quantum probe~\cite{cavazzoni2025frequencyestimationfrequencyjumps}? (ii) Can a proper sensing resource be defined to quantify enhancements in frequency estimation? (iii) Is it possible to link the enhancement in frequency estimation to the internal Hamiltonian structure?}}

In this Letter, we address all the above issues. Regarding the first question, we encode the unknown frequency into the dynamical state of a quantized electromagnetic field  with a nonlinear Hamiltonian. 
Once the frequency (unknown parameter) is dynamically encoded into a quantum state, we quantify the frequency precision limits using quantum estimation theory~\cite{helstrom1969quantum, paris2009quantum}. To address the second question, we define an energy cost figure of merit that quantifies enhancements in frequency estimation while balancing resource efficiency and estimation precision~\cite{liuzzo2018energy, rodriguez2024usefulness}. To address the third question, we examine our quantum sensing strategy in the broader context of quantum scrambling~\cite{garcia2023resource}, which describes how local quantum information disperses across the degrees of freedom in a quantum system~\cite{zanardi2001entanglement, zhou2017operator, khemani2018operator, chen2018operatorscramblingquantumchaos, zhou2023operatorgrowth, aleiner2016microscopic, harrow2021separation, nahum2018operator, keyserlingk2018operator, touil2024information, touil2020quantum, kobrin2024universalprotocolquantumenhancedsensing}. In contrast to prior quantum sensing studies that focus on dominant scrambling quantifiers~\cite{li2023improving}, we use the Wigner-Yanase skew information~\cite{wigner1963information, luo2003wigner, chen2005wigner, chen2023geometric, luo2019quantifying}, which enables us to quantify the degree of noncommutativity in the internal structure of the Hamiltonian~\cite{takagi2019skew, luo2012quantifying}. In particular, we show that higher-order nonlinearities enhance frequency sensitivity, with maximal sensitivity closely aligning with the maximum value of the Wigner-Yanase skew information. This observation highlights the connection between quantum scrambling, measured using the Wigner-Yanase skew information~\cite{luo2003wigner}, and the sensitivity of the probe to small changes in the unknown parameter.

\textit{Quantum metrological tools---} The uncertainty in estimating an unknown parameter $\omega$ encoded in a quantum state $\rho(\omega)$ obeys the quantum Cram\'{e}r-Rao theorem~\cite{cramer1999mathematical, rao1992breakthroughs, helstrom1967minimum, paris2009quantum}:
\begin{equation}
    \mathrm{Var}[\tilde{\omega}]\geq [M\mathcal{F}(\omega)]^{-1}\geq [MQ(\omega)]^{-1},\label{eq_QCRB}
\end{equation}
where $\mathrm{Var}[\tilde{\omega}]$ is the variance of a local unbiased estimator $\tilde{\omega}$ of the parameter $\omega$, $M$ is the number of measurement trials, $\mathcal{F}(\omega)$ is the classical Fisher information (CFI) with respect to the unknown parameter $\omega$, and $Q(\omega)$ is the quantum Fisher information (QFI) with respect to $\omega$. In the inequality of Eq.~\eqref{eq_QCRB}, the CFI is defined as $\mathcal{F}(\omega){:=}\sum_x p(x|\omega) \left[\partial_\omega \ln p(x|\omega)\right]^2$, where $\partial_\omega{:=}\frac{\partial}{\partial \omega}$, and $p(x|\omega){=}\mathrm{Tr}[\Pi_x \rho(\omega)]$ is the conditional probability distribution built from measurement statistics~\cite{cramer1999mathematical, rao1992breakthroughs}. Thus, the CFI sets the precision limits of estimating $\omega$ for a specific choice of positive-operator valued measure (POVM) $\{\Pi_x\}$ with measurement outcome $x$. The QFI is defined as the optimization over all possible POVMs, namely: $Q(\omega){:=}\max_{\{\Pi_x\}}\mathcal{F}(\omega)$. Alternatively, the QFI can also be defined as $Q(\omega){:=}\mathrm{Tr}[\partial_\omega \rho(\omega) L(\omega)]$. Here, $L(\omega)$ is the symmetric logarithmic derivative (SLD) operator that satisfies the equation $2\partial_\omega \rho(\omega){=}L(\omega) \rho(\omega){+}\rho(\omega) L(\omega)$, with the support of $L(\omega)$ providing the POVM basis that maximizes the CFI~\cite{paris2009quantum, liu2016quantum}. Throughout this work, we will only consider pure states $\rho(\omega){=}|\psi(\omega)\rangle \langle \psi(\omega)|$. In this specific case, the QFI simplifies to~\cite{statistical1994braunstein}: 
\begin{equation}
Q(\omega){=}4\mathbb{R}\text{e}\left[\langle \partial_\omega \psi(\omega) | \partial_\omega \psi(\omega) \rangle{-}|\langle \partial_\omega \psi(\omega) | \psi(\omega) \rangle|^2\right],
\end{equation}
which sets the ultimate precision limit for estimating the parameter $\omega$ and it quantifies the sensing capability of the quantum state $\rho(\omega)$ in the vicinity of $\omega$~\cite{degen2017quantum, sidhu2020geometric, giovannetti2004quantum, giovannetti2006quantum, giovannetti2011advances, paris2009quantum}.

\textit{The model---} We consider the Hamiltonian ($\hbar{=}1$),
\begin{equation}
    H=\omega H_0 + \beta H_1, \hspace{1cm} H_0:=a^\dagger a \label{eq_hamiltonian_probe}
\end{equation}
where $\omega$ is the unknown frequency that we aim to estimate, $\beta$ is a known parameter, $a$ ($a^\dagger$) is the annihilation (creation) operator obeying $[a{,}a^\dagger]{=}1$, and $H_1$ denotes a nonlinear Hamiltonian term. Throughout this work, we individually investigate three families of nonlinear $H_1$ terms, referred as: 
\begin{equation}
   H_1=\left\{
\begin{array}{ll}
      (a^\dagger{+}a)^s & \rightarrow \text{polynomial case,} \\
      a^{\dagger s}{+}a^s & \rightarrow \text{generalized squeezing case,} \\
      a^{\dagger s}a^s & \rightarrow \text{generalized Kerr case.} \\
\end{array} 
\right. \label{eq_family_of_nonlinearities}
\end{equation}
The unknown frequency will be encoded in $|\psi(t)\rangle{=}e^{-itH}|\psi(0)\rangle$ throughout the sensing strategy. For the sake of simplicity, we focus on coherent input fields with real amplitudes $|\psi(0)\rangle{=}|\alpha\rangle$, evolution times within the range of $\beta t{\leq}0.05$, and nonlinearity strength constrained to $\beta{\leq}\omega$. In general, we consider $\beta$ to be independent of $\omega$. For the specific case where $\beta{\propto}\omega$, see the Supplemental Material (SM)~\cite{SM}. 

To fairly claim \red{nonlinear-enhanced} frequency estimation due to $H_1$, we evaluate the QFI ratio $\mathfrak{R}(\omega){=}Q(\omega){/}Q_0(\omega)$ for a given sensing resource. Here, $Q(\omega)$ $[Q_0(\omega)]$ is the QFI with respect to $\omega$ for $\beta{>}0$ $[\beta{=}0]$, computed from an evolved initial coherent state $|\alpha\rangle$ $[|\alpha_0\rangle]$. For $\beta{=}0$, the QFI simplifies to $Q_0(\omega){=}4t^2\alpha_0^2$, which is independent of $\omega$. We define the sensing resource by imposing that both probes (for $\beta{>}0$ and $\beta{=}0$) have the same average energy $\langle\alpha_0| \omega a^\dagger a | \alpha_0 \rangle{=}\langle \alpha | H | \alpha \rangle$. Therefore, we obtain $\alpha_0^2{=}\langle \alpha | H | \alpha \rangle / \omega$, and then ($\alpha{\neq}0$):
\begin{equation}
\mathfrak{R}(\omega)=\frac{\omega Q(\omega)}{4t^2\langle\alpha{|}H{|}\alpha\rangle}.\label{eq_R}
\end{equation}
If the ratio $\mathfrak{R}(\omega){>}1$, then it indicates that \red{nonlinear-enhanced} sensing is achieved due to $H_1$ provided that both probes ($\beta{=}0$ and $\beta{>}0$) have the same energy on average. This energy constraint as resource implies that $\alpha_0{>}\alpha$. Hence, if $\mathfrak{R}(\omega){>}1$, adding a nonlinear term to the frequency estimation process is more energy-efficient when considering the initial number of excitations.

\vm{As a final note, one may wonder whether preparing the quantum state first through a nonlinear process generated by $\gamma H_1$, and subsequently encoding the unknown parameter $\omega$ via free evolution under $\omega H_0$, could lead to improved sensing performance. While in some cases this approach may indeed offer enhanced sensitivity compared to the current dynamical scenario---where the frequency is encoded via the combined Hamiltonian $H{=}\omega H_0{+}\beta H_1$---this advantage comes at the cost of requiring a very large nonlinearity ratio $\gamma{/}\beta$, with $\gamma$ chosen to correspond to a probe with the same energy as described above. Such high values of $\gamma$ lie far beyond the accessible regime $\beta{/}\omega$ considered in this work. A detailed analysis of this alternative scenario is presented in the SM~\cite{SM}.}

\textit{\blue{Polynomial case---}} This contribution can arise in nonlinear media as higher-order terms in the polarization $\boldsymbol{\mathrm{P}}~{\propto}~\chi^{(1)}\boldsymbol{\mathrm{E}}{+}\chi^{(2)}\boldsymbol{\mathrm{E}}{\cdot}\boldsymbol{\mathrm{E}}{+}\chi^{(3)}\boldsymbol{\mathrm{E}}{\cdot}\boldsymbol{\mathrm{E}}{\cdot}\boldsymbol{\mathrm{E}}{\ldots}$, where $\chi^{(i)}$ are the $i$th order nonlinear susceptibilities, and $\boldsymbol{\mathrm{E}}$ is the electric field. Since the interaction between the material and the radiation field is $-\boldsymbol{\mathrm{P}}{\cdot}\boldsymbol{\mathrm{E}}$, the nonlinear contributions to the interaction will include terms of the form $(a^\dagger{+}a)^s$, where $s$ is a positive integer. Anharmonic potentials of the form $V(x){\sim}x^2{+}\sum_sc_sx^s$, where $x{\sim}(a^\dagger{+}a)$ is the \textit{position} quadrature and $c_s{\in}\mathbb{R}$, can also lead to the same nonlinear contribution ${\sim}(a^\dagger{+}a)^s$. \blue{This polynomial contribution for $s=3$ has also been achieved experimentally in microwave superconducting systems \cite{hillmann2020universal, eriksson2024universal}.}

Two values of the exponents can be studied straightforwardly, namely: $s{=}1,2$. For $s{=}1$, the system is equivalent to a shifted harmonic oscillator with a displaced coherent amplitude $|\psi(t)\rangle_{s=1}{=}|\alpha e^{-i\omega t}{+}\frac{\beta}{\omega}(1{-}e^{-i\omega t})\rangle$~\cite{montenegro2014nonlinearity}. For the considered range of parameters $\{\beta{,}t{,}\omega\}$ in our analysis, $s{=}1$ results in \blue{negligible enhancements in frequency estimation}, i.e., $\mathfrak{R}{\sim}1$. See SM~\cite{SM} for details. For $s{=}2$, the additional quadratic term can be combined into a single term with a modified frequency. This leads to $|\psi(t)\rangle_{s=2}{=}|\alpha e^{-i\sqrt{\omega^2{+}4\beta\omega}t}\rangle$~\cite{qvarfort2022solvingquantumdynamicslie}. In this case, the QFI ratio satisfies $\mathfrak{R}{\leq}1$, see SM~\cite{SM} for details.

\begin{figure}[t]
\centering\includegraphics[width=\linewidth]{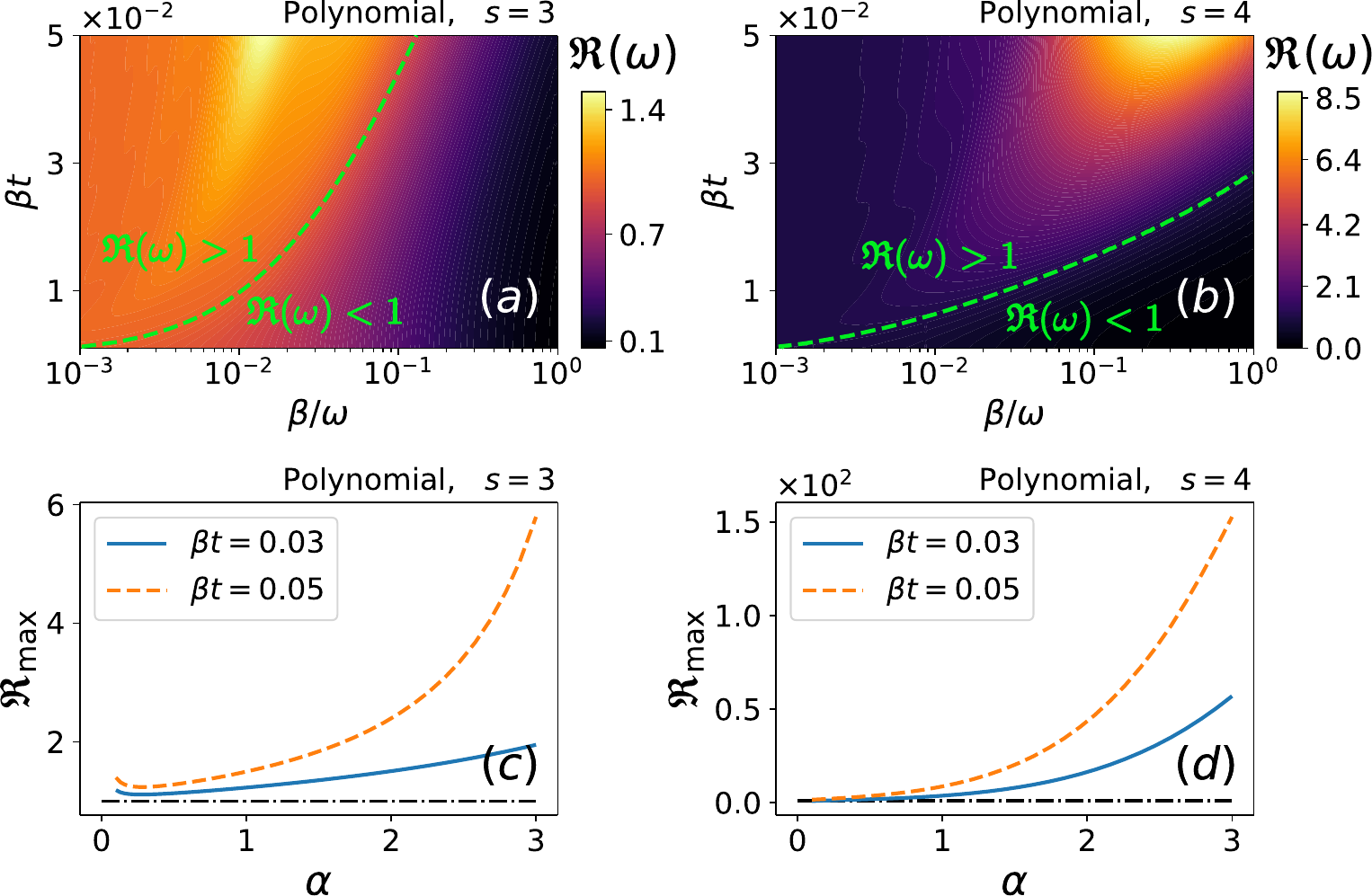}
    \caption{Polynomial case: Ratio $\mathfrak{R}(\omega)$ as functions of nonlinearity strength $\beta$ and time $t$. (a) $s{=}3$, (b) $s{=}4$. We use $\alpha{=}1$; Maximized ratio $\mathfrak{R}_\mathrm{max}$ as a function of $\alpha$ for different times $t$. (c) $s{=}3$, (d) $s{=}4$.}
    \label{fig_ratio_qfis}
\end{figure}

True nonlinearities arise when $s{\geq}3$. In Figs.~\ref{fig_ratio_qfis}(a)-(b), we plot the ratio $\mathfrak{R}(\omega)$ for the polynomial case as functions of the nonlinearity strength $\beta$ and time $t$ for $s{=}3$ and $s{=}4$, respectively. In Fig.~\ref{fig_ratio_qfis}(a), \red{nonlinear-enhanced} frequency estimation is achieved for several values of the nonlinear strength $\beta$ and time $t$. Specifically, when $\beta t{=}0.05$ and the nonlinearity is moderately weak $\beta{\sim}10^{-2}\omega$, the quantum enhancement ratio $\mathfrak{R}(\omega)$ approximates $1.5$. Note that as the nonlinear strength $\beta$ increases, the enhancement in frequency estimation decreases and eventually disappears. In Fig.~\ref{fig_ratio_qfis}(b), \red{nonlinear-enhanced} frequency estimation is amplified for several choices of $\beta$ and $t$, reaching a maximum enhancement ratio of $\mathfrak{R}(\omega){\sim}10$. However, in contrast to $s{=}3$ shown in Fig.~\ref{fig_ratio_qfis}(a), the maximum enhancement occurs at $\beta{\sim}10^{-1}\omega$, which is an order of magnitude higher than in the previous case.

A straightforward way to increase $\mathfrak{R}(\omega)$ is by increasing the initial coherent amplitude $\alpha$. To explore how $\mathfrak{R}(\omega)$ depends on $\alpha$, we calculate:
\begin{equation}
    \max_{0<\frac{\beta}{\omega}\leq 1}[\mathfrak{R}(\omega|\beta t)]:=\mathfrak{R}_\mathrm{max},
\end{equation}
which maximizes the value of $\mathfrak{R}(\omega|\beta t)$ over $\beta{/}\omega$ for a given $\alpha$ and time $\beta t$. In Fig.~\ref{fig_ratio_qfis}(c), we plot $\mathfrak{R}_\mathrm{max}$ for $s{=}3$ as a function of $\alpha$ for different times $t$. Two evident conclusion can be drawn from the figure: (i) \red{nonlinear-enhanced} frequency estimation ($\mathfrak{R}_\mathrm{max}{>}1$) can always be achieved for any coherent input $\alpha$, with $\mathfrak{R}_\mathrm{max}$ increasing as both time and the coherent amplitude grow; and (ii) the growth of $\mathfrak{R}_\mathrm{max}$ is super-linear in $\alpha$. Note that, in Fig.~\ref{fig_ratio_qfis}(c), a dip in $\mathfrak{R}_\mathrm{max}$ is shown for $\alpha{\leq}0.5$. This can be understood as $\mathfrak{R}(\omega){\gg}1$ when $\alpha{\rightarrow}0$. Similarly, in Fig.~\ref{fig_ratio_qfis}(d), we plot $\mathfrak{R}_\mathrm{max}$ for $s{=}4$ as a function of $\alpha$ for different times $t$. As the figure shows, a significant increase in the maximized ratio $\mathfrak{R}_\mathrm{max}$ is achieved as time and coherent input increase.

\begin{figure}[t]
\centering\includegraphics[width=\linewidth]{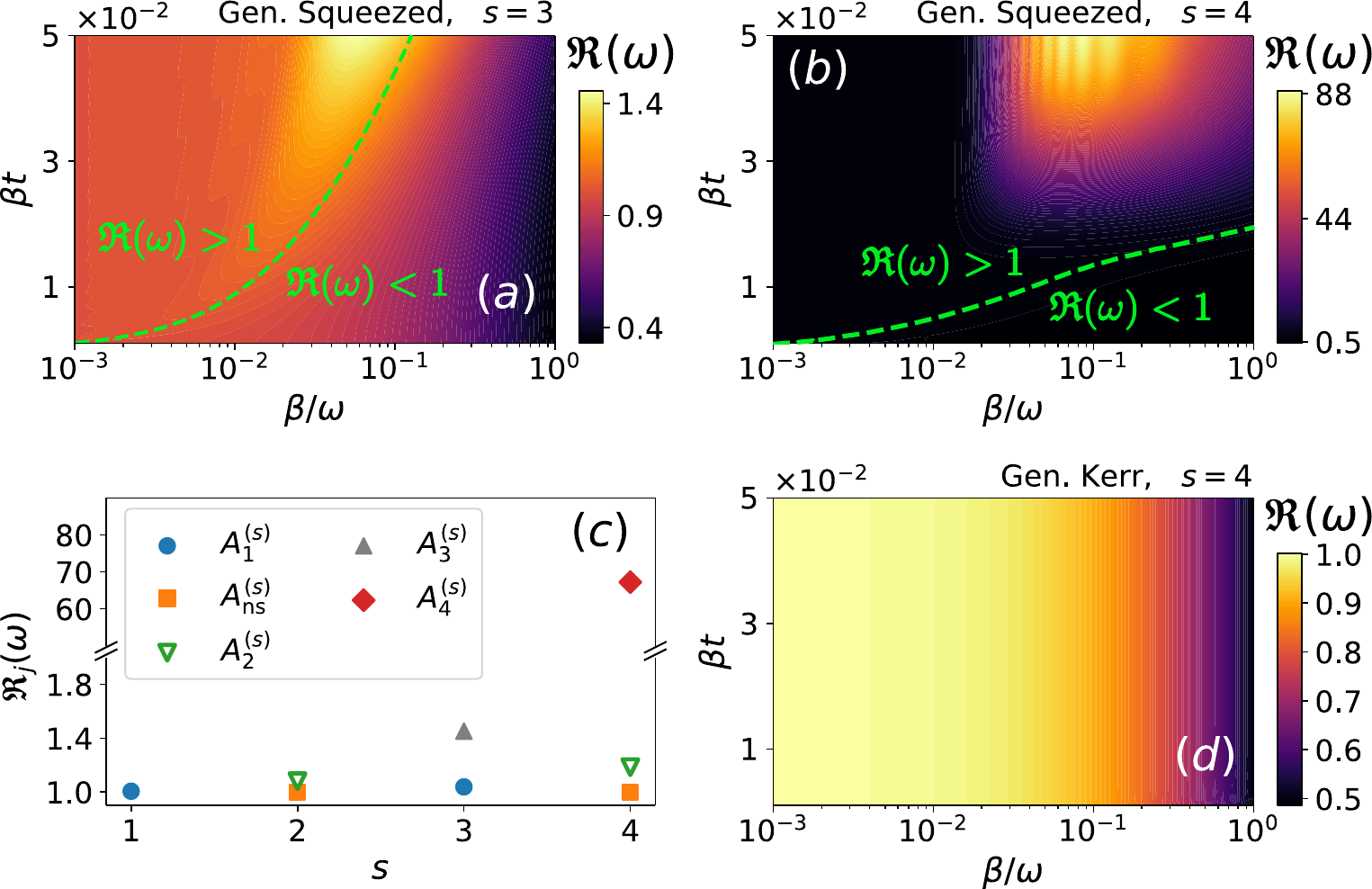}
    \caption{Generalized squeezing case: Ratio $\mathfrak{R}(\omega)$ as functions of nonlinearity strength $\beta$ and time $t$. (a) $s{=}3$, (b) $s{=}4$. (c) Ratio $\mathfrak{R}_j(\omega)$ as function of $s$ for several field excitation processes $A_j^{(s)}$. (d) Generalized Kerr case: Ratio $\mathfrak{R}(\omega)$ as functions of nonlinearity strength $\beta$ and time $t$ for $s{=}4$. We use $\alpha{=}1$.}
    \label{fig_sqz_qfis}
\end{figure}

\textit{\blue{Generalized squeezing case---}}  \blue{Generalized squeezing has been studied for decades in the context of quantum optics as a natural extension of standard second-order squeezing \cite{hong1985higher, hillery1984squeezing, d1987non, braunstein1987generalized, banaszek1997quantum, bencheikh2007triple}, and it has been recently attained to the third order in superconducting microwave systems \cite{chang2020observation}.} For $s{=}1$, the system reduces to the previously discussed polynomial case. For $s{=}2$, it corresponds to the well-known squeezing-driven case, where only a modest quantum enhancement in frequency estimation is observed within the parameter range considered, see SM~\cite{SM} for details. In Figs.~\ref{fig_sqz_qfis}(a)-(b), we plot $\mathfrak{R}(\omega)$ for the generalized squeezing case as functions of the nonlinearity $\beta$ and time $t$ for $s{=}3$ and $s{=}4$, respectively. As shown in Fig.~\ref{fig_sqz_qfis}(c), the generalized squeezing term for $s{=}3$ provides a quantum enhancement in frequency estimation comparable to the polynomial case. However, in Fig.~\ref{fig_sqz_qfis}(d), where $s{=}4$, significant frequency sensing enhancement is observed. To clarify this enhancement, we decompose the polynomial term in normal order as follows~\cite{louisell1990book,deepak2023generalexpansionnaturalpower}:
\begin{equation}
    (a^\dagger{+}a)^s{=}\sum_{k=0}^{\left[\frac{s}{2}\right]}\sum_{l=0}^{s-2k}\frac{s!}{2^kk!l!(s{-}2k{-}l)!}a^{\dagger s{-}2k{-}l}a^l.
\end{equation}
In particular, for $s{=}3$ and $s{=}4$ one gets:
\begin{eqnarray}
 (a^\dagger+ a)^3 &=& A_1^{(3)} + A_3^{(3)},\\
 (a^\dagger+ a)^4 &=& A_\mathrm{ns}^{(4)} + A_2^{(4)} + A_4^{(4)},
\end{eqnarray}
\blue{where the operators $A_{j}^{(s)}$ represents different field $j-$excitation process. We denoted with $A_j^{(s)}$ processes for which $A_j^{(s)}|n\rangle{\sim}|n{\pm}j\rangle$, and with $A_\mathrm{ns}^{(s)}$ processes for which $A_\mathrm{ns}^{(s)}|n\rangle{\sim}|n\rangle$ (where $|n\rangle$ is a Fock number state). Explicitly, for exponents $s{=}3$ one has: $ A_1^{(3)}{=}3[a^{\dagger 2}a{+}a^{\dagger}a^2{+}a^{\dagger}{+}a]$ and $A_3^{(3)}{=}a^{\dagger 3}{+}a^3$; and for $s{=}4$ one has: $A_2^{(4)}{=}4 a^{\dagger 3}a{+}4 a^\dagger a^{3}{+}6 a^{\dagger 2}{+}6 a^2$, $A_\mathrm{ns}^{(4)}{=}6[(a^\dagger a)^2{+}a^\dagger a]$, and $A_4^{(4)}{=}a^{\dagger 4}{+}a^4$.}
The observed improvement in frequency estimation in the generalized squeezing scenario is likely due to the increased role of higher-order field excitation processes, as indicated by the decomposition above. To quantify this, we calculate
\begin{equation}
    \mathfrak{R}_j(\omega) = \frac{\omega Q_j(\omega)}{4t^2\langle \alpha | \omega a^\dagger a {+} A_j^{(s)} | \alpha\rangle},
\end{equation}
where the ratio $\mathfrak{R}_j(\omega)$ quantifies possible \red{nonlinear-enhanced} frequency estimation due to individual $j$-field excitation processes. Consequently, $Q_j(\omega)$ is the QFI computed from a initial coherent state $|\alpha\rangle$ evolved under the action of the Hamiltonian $H_j{=}\omega a^\dagger a{+}A_j^{(s)}$, where $j{=}\text{ns}{,}1{,}2{,}\ldots$ and $s{=}1{,}2{,}\ldots$. Recall that $A_j^{(s)}$ accounts for individual $j$-field excitation processes derived from the decomposition of $(a^\dagger{+}a)^s$. To ensure a fair comparison and avoid the influence of multiplicative factors in the decomposition of $A_j^{(s)}$, we scale all individual terms $A_j^{(s)}$ to have the same average energy. For simplicity, this average energy is set to $\langle A_j^{(s)} \rangle{=}2\alpha$ and $\beta t{=}0.05$.

In Fig.~\ref{fig_sqz_qfis}(c), we plot the ratio $\mathfrak{R}_j(\omega)$ as a function of the exponent $s$ for several field excitation processes $A_j^{(s)}$. The figure shows that, for a given exponent $s$, \red{nonlinear-enhanced} frequency estimation consistently occurs for higher-order field excitation processes. In addition, number state processes $A_\mathrm{ns}^{(s)}$ offer no sensing advantage as for $s{=}2{,}4$ the ratio $\mathfrak{R}_j(\omega){=}1$. This can be explained as $[a^\dagger a{,}A_\mathrm{ns}^{(s)}]{=}0$ $\forall s$, and therefore, the evolved state is $|\psi(t)\rangle{=}e^{-i\beta t A_\mathrm{ns}^{(s)}}|\alpha e^{-i\frac{\omega}{\beta} t}\rangle$. As the unitary operator $e^{-i\beta t A_\mathrm{ns}^{(s)}}$ is independent of the unknown parameter to be estimated $\omega$, the QFI cannot increase by the addition of $A_\mathrm{ns}^{(s)}$ in the Hamiltonian provided that $[a^\dagger a{,}A_\mathrm{ns}^{(s)}]{=}0$.

\textit{\blue{Generalized Kerr case---}} \blue{The Kerr effect has also been extensively considered in quantum optics \cite{saleh2019fundamentals} and it emerges naturally in superconducting circuits due to the nonlinear inductance of a Josephson junction \cite{nigg2012black}.} By using $a^sa^\dagger{=}a^\dagger a^s{+}sa^{s{-}1}$ and $aa^{\dagger s}{=}sa^{\dagger (s-1)}{+}a^{\dagger s}a$, one can straightforwardly prove that $[a^\dagger a{,}a^{\dagger s}a^s]{=}0$. Thus no sensing benefit is expected from this additional term in the Hamiltonian. To see this, in Fig.~\ref{fig_sqz_qfis}(d), we plot the ratio $\mathfrak{R}(\omega)$ as functions of time $t$ and nonlinearity strength $\beta$ for $s{=}4$. As the figure shows, the ratio $\mathfrak{R}(\omega){\lesssim}1$, which implies no sensing advantage under this generalized Kerr scenario. Moreover, the ratio $\mathfrak{R}(\omega)$ decreases as $\beta{/}\omega$ increases. This situation occurs because we are comparing probes that have the same average energy. Specifically, $[a^\dagger a{,}a^{\dagger s} a^s]{=}0$ ensures that the numerator in the expression for $\mathfrak{R}(\omega)$ remains unchanged, whereas the denominator in $\mathfrak{R}(\omega)$ increases as $\beta$ increases when modified to $\omega \alpha^2{+}\beta \alpha^{2 s}$ for $\beta{\neq}0$. Thus, using a generalized Kerr term in the Hamiltonian does not provide any sensing advantage; in fact, a probe without nonlinearity is more energy-efficient.

\textit{Quantum scrambling---} Quantum scrambling has been studied across various fields, including quantum error correction~\cite{choi2020quantum}, machine learning~\cite{garcia2022quantifying, shen2020information, wu2021scrambling, holmes2021barren}, chemical reactions~\cite{zhang2024quantum}, and shadow tomography~\cite{garcia2021quantum, hu2022hamiltonian, mcginley2022quantifying, hu2023classical, bu2024classical}. Several metrics for measuring quantum scrambling have been proposed, such as operator entanglement entropy~\cite{zanardi2001entanglement, zhou2017operator}, average Pauli weight~\cite{khemani2018operator, zhou2019operator, chen2018operatorscramblingquantumchaos}, and the out-of-time-ordered correlator (OTOC)~\cite{zhou2023operatorgrowth, khemani2018operator, aleiner2016microscopic, roberts2015diagnosing, hosur2016chaos, harrow2021separation, nahum2018operator, keyserlingk2018operator}. The latter, OTOC, has even been experimentally demonstrated~\cite{landsman2019verified, mi2021information, li2017measuring, li2023improving}. Recently, a universal framework for information scrambling in open quantum systems was proposed~\cite{schuster2023operator}, along with its connection to quantum information thermodynamics~\cite{touil2020quantum, touil2024information} and a resource theory that encompasses both entanglement and magic scrambling mechanisms~\cite{garcia2023resource}. 

In the field of quantum sensing, quantum information scrambling has primarily been studied using OTOCs~\cite{kobrin2024universalprotocolquantumenhancedsensing, li2023improving}. Here, we explore an alternative approach by focusing on the Wigner-Yanase skew information~\cite{luo2003wigner}, which is defined as:
\begin{equation}
S(B, K){=}-\frac{1}{2} \mathrm{Tr} \left( [\sqrt{B}, K]^2 \right),~\label{eq_skew}
\end{equation}
which quantifies the degree of noncommutativity between a positive operator $B$ and a fixed Hermitian operator $K$. In our context, a natural choice for these operators is $B{=}|\psi(t)\rangle \langle\psi(t)|$, representing the time-evolved state, and $K{=}a^\dagger a $, the number operator. The Wigner-Yanase skew information, in this case, measures the extent to which the number operator fails to commute with the evolved state. In our case, $B$ is a pure state; thus, the Wigner-Yanase skew information simplifies to:
\begin{equation}
   S(|\psi(t)\rangle\langle\psi(t)|{,}a^\dagger a){=}\langle\psi(t)| (a^\dagger a)^2|\psi(t)\rangle{-}\langle\psi(t)|a^\dagger a|\psi(t)\rangle^2,~\label{eq_simplified_skew}
\end{equation}
\begin{figure}[t]
    \centering
    \includegraphics[width=\linewidth]{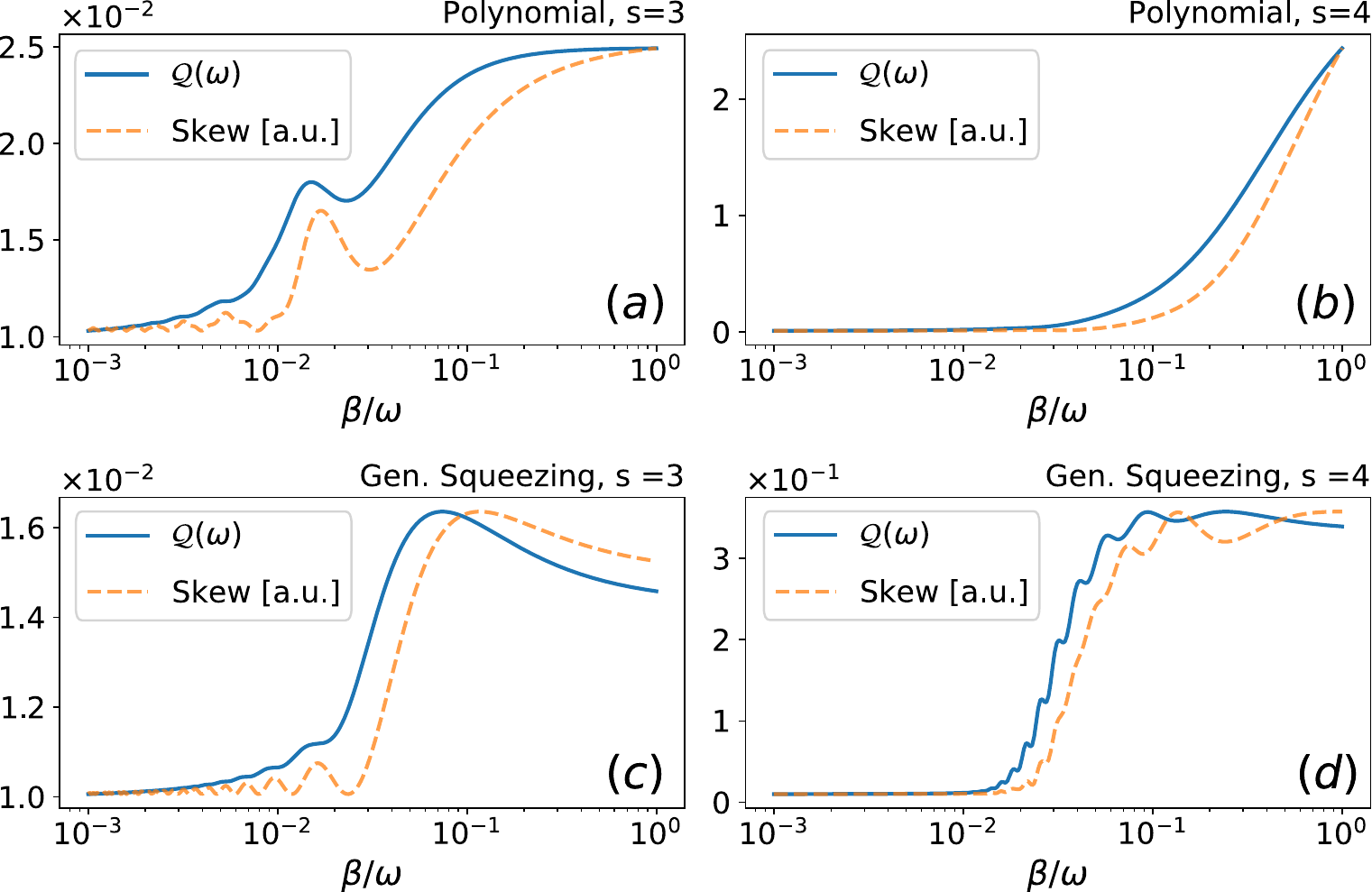}
    \caption{Skew information and the QFI as functions of the nonlinearity $\beta$ for $s{=}3{,}4$ at a fixed time $\beta t{=}0.05$. The polynomial case is shown in panels (a) and (b); The generalized squeezing case is shown in panels (c) and (d).}
    \label{fig_skew}
\end{figure}
\noindent which is the variance of the observable $K{=}a^\dagger a$ in the state $|\psi(t)\rangle$. Note that, if $[H_0{,}H_1]{=}0$, \red{then $S(|\psi(t)\rangle\langle\psi(t)|{,}a^\dagger a){=}|\alpha|^2$.} In Fig.~\eqref{fig_skew}, we plot the skew information and the QFI as functions of the nonlinearity $\beta$ for the polynomial and generalized squeezing scenarios for $s{=}3{,}4$ at a fixed time $\beta t{=}0.05$. For clarity, the skew information is presented in arbitrary units [a. u.]. As the figure shows, the skew information fully captures the behavior of the QFI. Notably, the maximum value of the skew information (i.e., where $H_0$ and $H_1$ are maximally noncommutative) occurs near the maximum value of the QFI (i.e., where the quantum probe is most sensitive to $\omega$). 
\vm{Note that for the family of states $\rho(\omega){=}e^{-i t \omega H_0} \rho(0) e^{i t \omega H_0}$, the QFI $Q(\omega){=}4t^2(\Delta H_0)^2$~\cite{paris2009quantum} exactly coincides with the Wigner–Yanase skew information. A formal proof of this equivalence can be found in Ref.~\cite{luo2004wigneryanaseskewinformationvs}. However, in our scenario, the unknown frequency is encoded via $H{=}\omega H_0{+}\beta H_1$. As a result, a slight quantitative discrepancy arises between the QFI and the Wigner–Yanase skew information. This is due to their fundamentally different mathematical natures. Indeed, the QFI defines a strict Riemannian metric on the space of quantum states, capturing the infinitesimal statistical distinguishability between nearby states~\cite{braunstein1994statistical}. In contrast, the Wigner–Yanase skew information given in Eq.~\eqref{eq_skew} is not a metric. Rather, it is a QFI-like quantity defined by $4\mathrm{Tr}\left(\partial_\omega\sqrt{\rho(\omega)}\right)^2,
$ which captures sensitivity to parameter changes without relying on the SLD~\cite{luo2003wigner}. Generalizations of the Wigner–Yanase skew information into a metric form have been explored using the Morozova–Chentsov functions, leading to the concept of metric-adjusted skew information~\cite{yang2022generalized, hansen2008metric}. However,} given that the skew information in Eq.~\eqref{eq_simplified_skew} is computationally more tractable than the QFI, it can be used to optimize a nonlinear quantum probe with the additional Hamiltonian $H_1$ by decomposing it into specific transition operators $A_j^{(s)}$, such that (possible) time-dependent functions $f_s(\beta_j{,}t)$ can be optimized in $H{=}\omega a^\dagger a{+}\sum_j f_s(\beta_j{,}t) A_j^{(s)}$ for maximizing $\mathcal{R}(\omega)$. This optimization procedure makes \red{nonlinear-enhanced} frequency sensing more efficient for complex interactions where these processes dominate.

\textit{Conclusions---} In this Letter, we have shown that \red{nonlinear-enhanced} frequency sensing can be achieved by efficiently encoding the frequency of a quantized electromagnetic field into a nonlinear quantum probe. The introduction of nonlinearities into the quantum probe facilitates the distribution of local quantum information across a larger Hilbert space, a process referred to as quantum scrambling. To make a fair comparison between probes with and without nonlinearities, we introduced a figure of merit that ensures both probes have the same average energy. To show the generality of our findings, we examined three distinct families of nonlinear contributions. We showed that higher-order nonlinearities lead to an enhanced sensing capacity, especially in regions where the noncommutativity between the number state observable and the nonlinear contribution is maximal. As a result, we established a connection between quantum scrambling, quantified by the Wigner-Yanase skew information, and the probe's sensitivity to slight variations in an unknown parameter, quantified by the quantum Fisher information. \textcolor{black}{Our results show that improved frequency estimation may be obtained under realistic conditions (nonlinear coupling $\beta/\omega \simeq 0.01-0.1$, interaction time $\beta t \leq 0.05$ and coherent amplitude $\alpha\simeq 1-5$) making them suitable for applications in enhancing magnetometry with superconducting quantum circuits and improving stability in superconducting cavity clocks.}

\textit{Acknowledgements---} V.M. thanks support from the National Natural Science Foundation of China Grants No. 12374482 and No. W2432005. MGAP is partially supported by EU 
and MIUR through the project PRIN22-2022T25TR3-RISQUE.

\bibliography{Scrambling}

\clearpage
\onecolumngrid
\widetext
\begin{center}
\textbf{\large Supplemental Material: Enhanced quantum frequency estimation by nonlinear scrambling}
\end{center}

\begin{center}
    Victor Montenegro$^{1,2,3}$, Sara Dornetti$^{4}$, Alessandro Ferraro$^{4}$, and Matteo G. A. Paris$^{4}$\\
    \vspace{0.2cm}
    \textit{$^1$\small{College of Computing and Mathematical Sciences, Department of Applied Mathematics and Sciences, Khalifa University of Science and Technology, 127788 Abu Dhabi, United Arab Emirates}}\\
    \textit{$^2$\small{Institute of Fundamental and Frontier Sciences, University of Electronic Science and Technology of China, Chengdu 611731, China.}}\\
    \textit{$^3$\small{Key Laboratory of Quantum Physics and Photonic Quantum Information, Ministry of Education, University of Electronic Science and Technology of China, Chengdu 611731, China.}}\\
    \textit{$^4$\small{Quantum Technology Lab $\&$ Applied Quantum Mechanics Group, Dipartimento di Fisica ``Aldo Pontremoli'', Universit\`a degli Studi di Milano, I-20133 Milano, Italia}}
    
\end{center}

\setcounter{equation}{0}
\setcounter{figure}{0}
\setcounter{table}{0}
\setcounter{page}{1}
\makeatletter
\renewcommand{\theequation}{S\arabic{equation}}
\renewcommand{\thefigure}{S\arabic{figure}}

\textbf{Outline:}

\begin{enumerate}
    \item[\textbf{I.}] Frequency Estimation with a Specific Nonlinearity Contribution: Polynomial Case
    \item[\textbf{II.}] Frequency Estimation with a General Nonlinearity Contribution: Polynomial Case for s=1,2
    \item[\textbf{III.}] Frequency Estimation with a General Nonlinearity Contribution: Generalized Squeezing Case for s=2
    \item[\textbf{IV.}] Frequency Estimation with a General Nonlinearity Contribution: Heterodyne Sensing Performance
    \item[\textbf{V.}] Alternative Two-Step Encoding Strategy
\end{enumerate}

\section{I. Frequency Estimation with a Specific Nonlinearity Contribution: Polynomial Case}

There is a particular situation in the polynomial case, in which the QFI can be computed analytically using coherent states. The interaction term in this case is slightly different from the one employed in the main text: assuming that the nonlinearity strength increases linearly with the frequency, the Hamiltonian takes the form: 

\begin{equation}
    H = \omega a^\dagger a+ \omega\beta (a+a^\dagger)^s \equiv \omega (a^\dagger a+ \beta G_s) \equiv \omega \mathcal{G}_s.
\end{equation}

The QFI for a coherent state that evolves according to $U(t)=e^{-iHt}$ has the simple form 
 $Q_s=4t^2[\langle\alpha|\Delta\mathcal{G}_s^2|\alpha\rangle]$. In order to compute $Q_s$, the idea is to rewrite $\mathcal{G}_s$ and $\mathcal{G}_s^2$ in normal order, exploiting identities that can be found starting from the commutation relation $[a,a^\dagger]=1$, i.e.: $a^ja^{\dagger}=a^\dagger a^j+ja^{j-1}$,  $aa^{\dagger j}=ja^{\dagger j-1}+a^{\dagger j}a$ and, in general: 

\begin{equation} 
    a^na^{\dagger m}=\sum_{j=0}^{\mathrm{min}[m,n]} j!\binom{n}{j}\binom{m}{j}a^{\dagger m-j}a^{n-j}.
\end{equation}
The first thing to do is to order $G_s$: 

\begin{equation} \label{Eq:G_s}
   G_s=(a+a^\dagger)^s=\sum_{k=0}^{[s/2]}\sum_{l=0}^{s-2k}\frac{s!}{2^k k! l! (s-2k-l)!}a^{\dagger s-2k-l }a^{l}.
\end{equation}

\noindent A complete proof of Eq. (\ref{Eq:G_s}) can be found in~\cite{deepak2023generalexpansionnaturalpower}. Now that $\mathcal{G}_s$ is taken care of, it is possible to order the second term $\mathcal{G}_s^2=a^\dagger a+ a^{\dagger 2}a^2 + \beta a^\dagger aG_s+ \beta G_sa^\dagger a+\beta^2G_sG_s $ using the relations above:
\begin{equation}
    a^\dagger aG_s=\sum_{k=0}^{[s/2]}\sum_{l=0}^{s-2k}\sum_{j=0}^{\text{min}[1,s-2k-l]}j!\binom{1}{j}\binom{s-2k-l}{j}\frac{s!}{2^k k! l! (s-2k-l)!}a^{\dagger s-2k-l-j+1 }a^{l-j+1},
\end{equation}

\begin{equation}
    G_s a^\dagger a=\sum_{k=0}^{[s/2]}\sum_{l=0}^{s-2k}\sum_{j=0}^{\text{min}[1,l]}j!\binom{1}{j}\binom{l}{j}\frac{s!}{2^k k! l! (s-2k-l)!}a^{\dagger s-2k-l-j+1 }a^{l-j+1},
\end{equation}

\begin{equation} 
   G_s G_s=G_{2s}=\sum_{k=0}^{s}\sum_{l=0}^{2s-2k}\frac{(2s)!}{2^k k! l! (2s-2k-l)!}a^{\dagger 2s-2k-l }a^{l}.
\end{equation}

In order to make clear the dependence of the QFI on the coherent state, it is useful to rewrite the complex number $\alpha$ in its exponential form $\alpha=re^{i\phi}$, where $r=|\alpha|\geq 0$ and  $0\leq\phi < 2\pi$. 

\begin{align*}
    Q_1(r,\phi, \beta)&=4 t^2 [ (r^2 + 2 r \beta + \beta^2) -  r \beta \phi^2 + o(\phi^3)] \\
    Q_2(r,\phi, \beta)&= 4t^2[ (r^2 + 8 r^2 \beta + 2 \beta^2 + 16 r^2 \beta^2) - 8 r^2 \beta (1 + 2 \beta) \phi^2 + o(\phi^3) ] \\
    Q_3(r,\phi, \beta)&=  4t^2[ (r^2 + 6 r \beta + 24 r^3 \beta + 15 \beta^2 + 144 r^2 \beta^2 + 144 r^4 \beta^2) - 3 r \beta (1 + 12 r^2 + 48 r \beta + 96 r^3 \beta) \phi^2 + o(\phi^3)] \\
    Q_4(r,\phi, \beta)&=4t^2[ (r^2 + 48 r^2 \beta + 64 r^4 \beta + 96 \beta^2 + 1536 r^2 \beta^2 + 2688 r^4 \beta^2 + 1024 r^6 \beta^2)+ \\
    & - 16 r^2 \beta (3 + 8 r^2 + 96 \beta + 336 r^2 \beta + 192 r^4 \beta)\phi^2+ o(\phi^3)]
\end{align*}   

\noindent The series expansions for $s = 1, 2, 3, 4$ show that there is a local maximum of the QFI when the phase is null. Consequently, the choice of a real $\alpha$ is not merely a simplification. From this moment on, we will always assume $\alpha$ to be real. Under this assumption, the expression for $\mathcal{G}_s^2$ can be further simplified, since the expectation values of $G_s a^\dagger a$ and $a^\dagger a G_s$ are equal. This equality follows from the fact that $(G_s a^\dagger a)^\dagger = a^\dagger a G_s$, which holds due to the hermiticity of $G_s$. Thus, the expectation value of $\mathcal{G}_s^2$ becomes:

\begin{equation}
    \langle\alpha| \mathcal{G}_s^2 |\alpha\rangle =\langle \alpha| a^\dagger a + a^{\dagger 2} a^2+ 2\beta a^\dagger aG_s +\beta^2G_{2s}|\alpha\rangle,
\end{equation}

\noindent and the final expression of the QFI is: 

\begin{equation}
    Q_s=4t^2[\alpha^2 + 2 \beta\langle a^\dagger a G_s\rangle+\beta^2\langle G_{2s} \rangle -2\beta\alpha^2\langle G_s \rangle -\beta^2\langle G_s\rangle^2].
\end{equation}

As in the main text, we impose that the probes ($\beta=0$ and $\beta>0$) have the same average energy $\langle \alpha_0 |\omega a^\dagger a |\alpha_0\rangle=\langle \alpha| \omega \mathcal{G}_s | \alpha\rangle$. Consequently, we obtain $
\alpha_0=\sqrt{\alpha^2+\beta \langle G_s \rangle}$. The expansion of the ratio $\mathfrak{R}_s(\alpha,\beta)=\frac{Q_s(\alpha,\beta)}{Q_0(\alpha_0)}$ for $s=1,2,3,4$ is reported below, both for small coherent amplitude $\alpha$ and small nonlinearity strength $\beta$:

\noindent
\begin{subequations}
\begin{minipage}{0.50\textwidth}
\begin{equation} \label{Eq:R_expansion_alpha}
    \begin{aligned}
        & \mathfrak{R}_1(\alpha,\beta)=\frac{\beta}{2\alpha}+\frac{3}{4}+\frac{\alpha}{8\beta}+o(\alpha^2)\\
        & \mathfrak{R}_2(\alpha,\beta)=2\beta +\left(6+\frac{1}{\beta}+8\beta \right)\alpha^2+o(\alpha^3)\\
        & \mathfrak{R}_3(\alpha,\beta)=\frac{5\beta}{2\alpha}+\frac{7}{12}+\left(\frac{5}{72\beta}+\frac{62\beta}{3}\right)\alpha+o(\alpha^2)\\
        & \mathfrak{R}_4(\alpha,\beta)=32\beta+\frac{1}{3}\left(16+\frac{1}{\beta}+768\beta\right)\alpha^2+o(\alpha^3)
    \end{aligned}
\end{equation}
\end{minipage}
\hfill
\begin{minipage}{0.50\textwidth}
\begin{equation} \label{Eq:R_expansion_beta}
    \begin{aligned}
        & \mathfrak{R}_1(\alpha,\beta)=1+\frac{\beta^2}{\alpha^2}+o(\beta^3)\\
        & \mathfrak{R}_2(\alpha,\beta)=1+\left(4-\frac{1}{\alpha^2}\right)\beta+o(\beta^2)\\
        & \mathfrak{R}_3(\alpha,\beta)=1+16\alpha\beta+o(\beta^2)\\
        & \mathfrak{R}_4(\alpha,\beta)=1+\left(24-\frac{3}{\alpha^2}+48\alpha^2\right)\beta+o(\beta^2).
    \end{aligned}
\end{equation}
\end{minipage}
\end{subequations}

\bigskip

\noindent The expansion with respect to small $\alpha$ in Eq. (\ref{Eq:R_expansion_alpha}) shows that for $s=1,3$, the ratio can be made arbitrarily large with an appropriate choice of $\beta\gg\alpha$. However, this property does not hold for 
$s=2,4$. The expansion with respect to small $\beta$ in Eq. (\ref{Eq:R_expansion_beta}), shows that the ratio is equal to $1$ plus a quantity that is always positive for $s=1,3$, but can be negative for $s=2,4$. As a result, in the latter case, enhancement in frequency estimation is not universally guaranteed.

\section{II. Frequency Estimation with a General Nonlinearity Contribution: Polynomial case for s=1,2}

\begin{figure}
    \centering
    \includegraphics[width=0.85\linewidth]{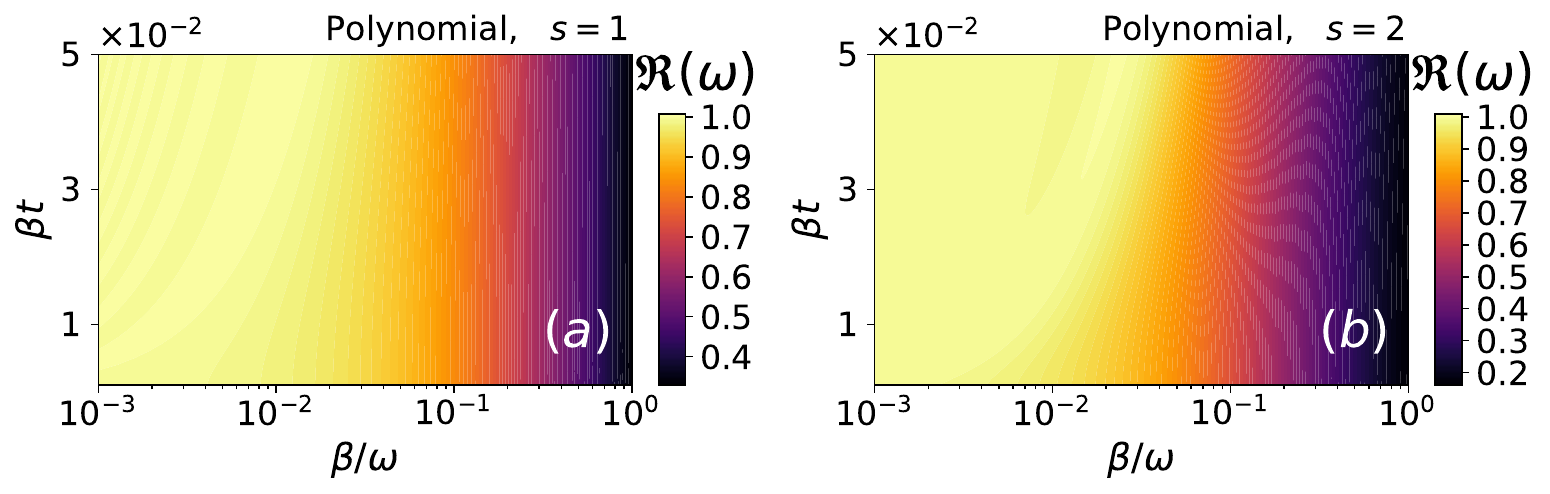}
    \caption{QFI ratio $\mathfrak{R}(\omega)$ for the polynomial case as functions of time $\beta t$ and the nonlinearity strength $\beta/\omega$. No \red{enhancements in} frequency estimation is reported, namely: $\mathfrak{R}(\omega) \leq 1$. (a) $s{=}1$, (b) $s{=}2$.}
    \label{fig_sm_polynomial_s12}
\end{figure}

Two straightforward polynomial scenarios were mentioned in the main text, namely: when the exponents are $s=1$ and $s=2$. On the one hand, for the polynomial case $s=1$, the quantum probe can be interpreted as a shift in the equilibrium position of the quantum harmonic oscillator (or as time-independent parametric external driving). Indeed, the Hamiltonian for the polynomial nonlinear case is given by $H = \omega a^\dagger a + \beta (a^\dagger + a)$, which upon switching to the position $x$ and momentum $p$ quadratures,
\begin{equation}
x = \sqrt{\frac{\hbar}{2\mu\omega}} (a + a^\dagger), \hspace{2cm} p = i\sqrt{\frac{\mu\hbar\omega}{2}} (a^\dagger - a), \label{eq_quadratures}
\end{equation}
the Hamiltonian becomes $H \sim \frac{p^2}{2\mu} + \frac{\mu\omega^2}{2}\left(x + x_0\right)^2$, where $x_0 = \frac{\beta}{\mu\omega^2}\sqrt{\frac{2\mu\omega}{\hbar}}$ is the shift of the oscillator's equilibrium position, with $\mu$ being the effective mass of the oscillator. In this scenario, $s{=}1$, the temporal unitary operator can be found straightforwardly as~\cite{montenegro2014nonlinearity}:  
\begin{equation}  
    U(t) = D\left[\frac{\beta}{\omega} (1 - e^{-i\omega t})\right] e^{-i \omega a^\dagger a t},
\end{equation}  
where $D[z] = \exp(z a^\dagger - z^* a)$ is the displacement operator. Therefore, an initial coherent state $|\alpha\rangle$ evolves as:  
\begin{equation}  
    |\psi(t)\rangle_{s=1} = \left|\alpha e^{-i\omega t} + \frac{\beta}{\omega} \left(1 - e^{-i\omega t}\right)\right\rangle, \label{eq_state_polynomial_s1}
\end{equation}  
which is the expression shown in the main text. Using Eq.~\eqref{eq_state_polynomial_s1}, the quantum Fisher information (QFI) $Q(\omega)$ with respect to $\omega$ can be directly evaluated. Consequently, the QFI ratio $\mathfrak{R}(\omega)$ is:
\begin{equation}
\mathfrak{R}(\omega)=\frac{\omega Q(\omega)}{4t^2\langle\alpha{|}H{|}\alpha\rangle}=\frac{2 \alpha \beta t^2 \omega^3 + \alpha^2 t^2 \omega^4 + \beta^2 (2 + t^2 \omega^2) - 2 \beta^2 \cos(t \omega) - 2 \beta t \omega (\beta + \alpha \omega) \sin(t \omega)}{t^2 \omega^3 (2 \alpha \beta + \alpha^2 \omega)}.
\end{equation}
In Fig.~\ref{fig_sm_polynomial_s12}(a), we plot the QFI ratio $\mathfrak{R}(\omega)$ for the polynomial case $s{=}1$ as functions of time $\beta t$ and the nonlinearity strength $\beta/\omega$. As the figure shows, within the range of parameters considered in this work, the QFI ratio indicates no \red{enhancements in} frequency estimation, namely: $\mathfrak{R}(\omega) \leq 1$. The latter corresponds to the claim presented in the main text. 

On the other hand, for the polynomial case $s=2$, the quantum probe can be interpreted as squeezing induced by modulation of the oscillator's frequency. Here, the Hamiltonian is $H = \omega a^\dagger a + \beta (a^\dagger + a)^2 \sim (\omega + 2\beta)a^\dagger a + \beta a^{\dagger 2} + \beta a^2$. The above Hamiltonian can be readily diagonalized using a Bogoliubov transformation or by rewriting the Hamiltonian in terms of the position $x$ and momentum $p$ [see Eqs.~\eqref{eq_quadratures}]. Hence:
\begin{equation}
H = \frac{p^2}{2\mu} + \frac{1}{2}\mu\omega^2 x^2 + \frac{1}{2}\mu\omega'^2 x^2.
\end{equation}
In this formulation, the polynomial nonlinear term $\beta (a^\dagger + a)^2 = \frac{\hbar w'^2}{4\omega} (a^\dagger + a)^2$ translates directly into an additional quadratic potential proportional to $x^2$. By combining the coefficients of $x^2$, the effective frequency of the oscillator is modified. The resulting diagonalized Hamiltonian becomes:
\begin{equation}
H \sim \frac{p^2}{2\mu} + \frac{1}{2}\mu\omega_{\text{eff}}^2 x^2 = \hbar \omega_{\text{eff}} b^\dagger b = \hbar \sqrt{\omega^2 + 4\omega\beta} b^\dagger b,\label{eq_sm_bogoliubov}
\end{equation}
where $x=\sqrt{\frac{\hbar}{2\mu\omega_{\text{eff}}}} (b + b^\dagger), p =i\sqrt{\frac{\mu\hbar\omega_{\text{eff}}}{2}} (b^\dagger - b)$. Eq.~\eqref{eq_sm_bogoliubov} coincides with the one presented in the main text.

Solutions for the temporal unitary operator can be obtained using a Lie algebraic approach~\cite{qvarfort2022solvingquantumdynamicslie}. However, a closed analytical form for the evolution of an initial coherent state under $H = \omega a^\dagger a + \beta (a^\dagger + a)^2$ is not attainable. In Fig.~\ref{fig_sm_polynomial_s12}(b), we numerically evaluate the QFI ratio $\mathfrak{R}(\omega)$ for the polynomial case $s{=}2$ as functions of time $\beta t$ and the nonlinearity strength $\beta/\omega$. As seen in the figure, within the range of parameters considered in this work, the QFI ratio indicates no \red{enhancements in} frequency estimation, namely $\mathfrak{R}(\omega) \leq 1$. This supports our claim in the main text for this specific case. 

\section{III. Frequency Estimation with a General Nonlinearity Contribution: Generalized squeezing case for s=2}

\begin{figure}
    \centering
    \includegraphics[width=0.49\linewidth]{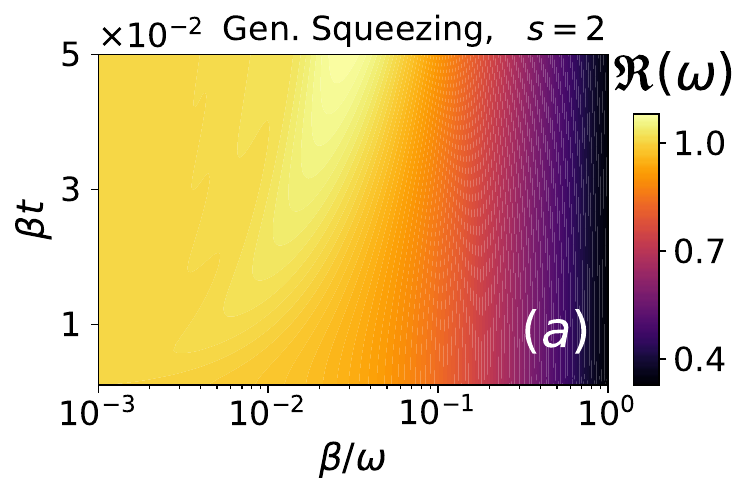}
    \caption{QFI ratio $\mathfrak{R}(\omega)$ for the generalized squeezing case $s{=}2$ as functions of time $\beta t$ and the nonlinearity strength $\beta/\omega$.}
    \label{fig_sm_squeezing_s2}
\end{figure}

In contrast to the previous polynomial scenario, for the generalized squeezing case $H = \omega a^\dagger a + \beta a^{\dagger 2} + \beta a^2$, the number state contribution depends only on the frequency we aim to estimate and is not affected by additional nonlinearities $\beta$. This implies that: (i) the Hamiltonian for the generalized squeezing case $s = 2$ cannot be rewritten as being proportional to $\sim b^\dagger b$; and (ii) increasing the nonlinearity only enhances the two-field excitation process. In contrast, for the polynomial case, both the number state subspace and the two-field excitation process are modified as the nonlinearity increases. In Fig.~\ref{fig_sm_squeezing_s2}(a), we numerically compute the QFI ratio $\mathfrak{R}(\omega)$ for the generalized squeezing case $s{=}2$ as functions of time $\beta t$ and the nonlinearity strength $\beta/\omega$. As shown in the figure, within the range of parameters considered in this work, the QFI ratio shows negligible \red{nonlinear-enhanced} frequency estimation $\mathfrak{R}(\omega)\sim 1$. Note that, for the case $s = 2$, by increasing the time $\beta t$, one can achieve higher values of $\mathfrak{R}(\omega)$ (not shown in the figure). Nonetheless, these values would fall outside current experimental capabilities. Therefore, we restrict ourselves to exploring the set of parameters within experimental reach, namely $\beta t \leq 0.05$ and $\beta \leq \omega$. Further studies involving longer times and stronger nonlinearities could be explored in future work.

\section{IV. Frequency Estimation with a General Nonlinearity Contribution: Heterodyne sensing performance}

Any sensing protocol relies on performing measurements on the probe to extract information about the parameters of interest. The outcomes of these measurements are used to construct probability distributions, which are then used to build the classical Fisher information (CFI). As discussed in the ``Quantum Metrological Tools" section, the symmetric logarithmic derivatives (SLDs) $L(\omega)$ define the optimal measurement basis required to achieve the ultimate sensing precision, quantified by the quantum Fisher information (QFI) $Q(\omega) := \mathrm{Tr}[\partial_\omega \rho(\omega) L(\omega)]$. However, while the optimal measurement (based on the eigenstates of the SLDs) is theoretically well-defined~\cite{paris2009quantum}, in practice, such measurements are often highly correlated and challenging to implement. Therefore, it is far more informative to establish the sensing precision for a given feasible measurement basis, which in the case of the electromagnetic field is typically limited to photocounting, homodyne detection, and heterodyne detection. Given the large Hilbert space due to the presence of $H_1$, heterodyne detection can be efficiently simulated using probability distributions 
\begin{equation}
p(\Upsilon{|}\omega){=}\frac{1}{\pi}\text{Tr}[|\Upsilon\rangle\langle\Upsilon|\psi(t)\rangle\langle\psi(t)|],
\end{equation}
where $|\Upsilon\rangle$ a coherent state with complex amplitude. Hence, the CFI, see ``Quantum Metrological Tools" section, is:
\begin{equation}
    \mathcal{F}(\omega)^\mathrm{Heterodyne}=\int d^2\Upsilon  p(\Upsilon|\omega) \left[\partial_\omega \ln p(\Upsilon|\omega)\right]^2.
\end{equation}

\begin{figure}
    \centering
    \includegraphics[width=\linewidth]{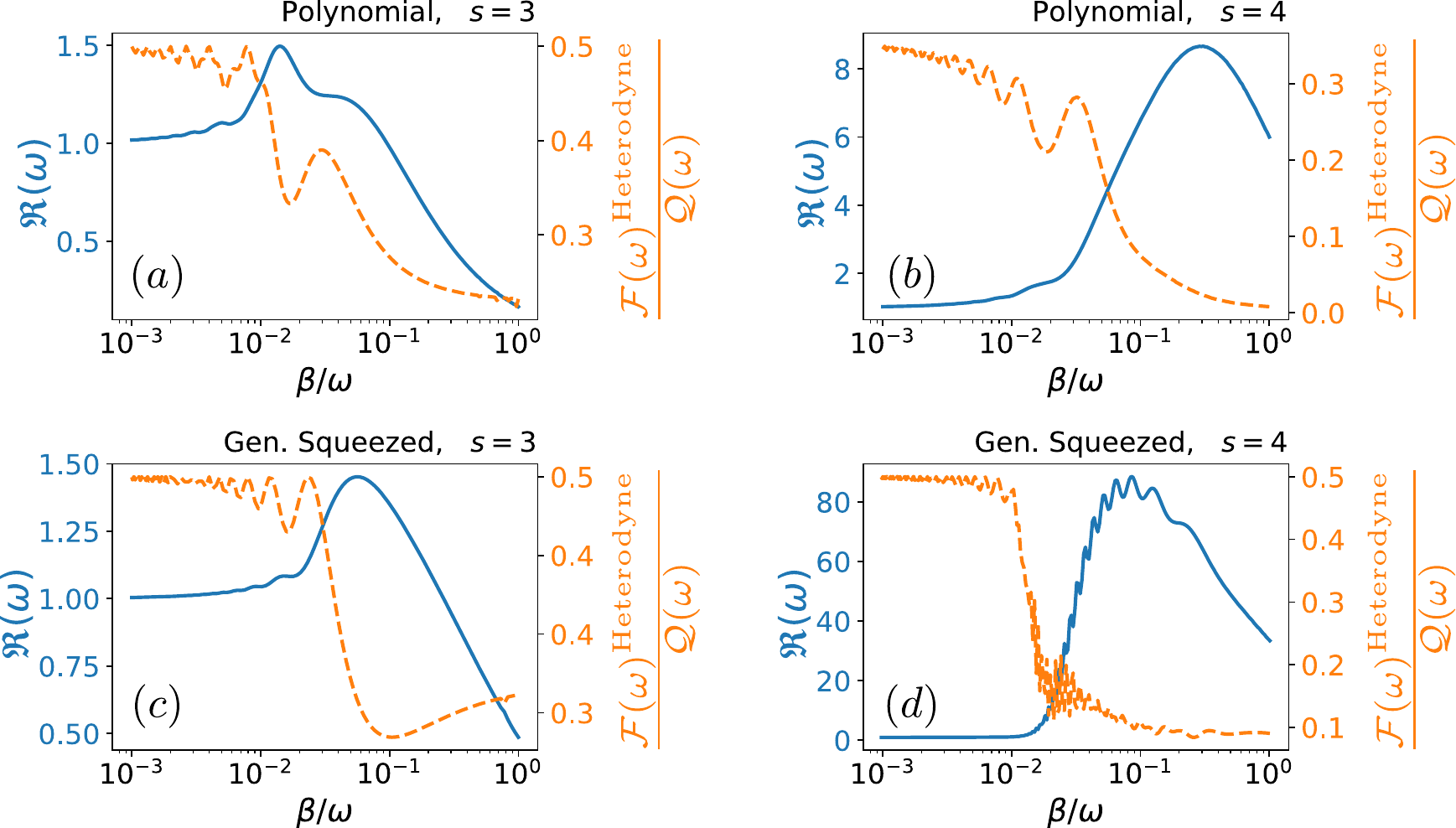}
    \caption{Heterodyne performance ratio $\mathcal{F}(\omega)^\mathrm{Heterodyne}/Q(\omega)$ as a function of the nonlinearity strength $\beta$. Panels (a) and (b), polynomial case for $s=3$ and $s=4$. Panels (c) and (d), generalized squeezing case for $s=3$ and $s=4$. For reference, we show the QFI ratio $\mathcal{R}(\omega)$. We set $\beta t = 0.05$ and the initial coherent amplitude $\alpha=1$.}
    \label{fig_heterodyne}
\end{figure}

To assess the sensing performance of heterodyne detection, we calculate the ratio between the CFI obtained from heterodyne detection and the QFI: $\mathcal{F}(\omega)^\mathrm{Heterodyne}/Q(\omega)$. In Fig.~\ref{fig_heterodyne}, we plot $\mathcal{F}(\omega)^\mathrm{Heterodyne}/Q(\omega)$ as a function of the nonlinearity strength $\beta$ for both the polynomial and generalized squeezing scenarios. For reference, we also include the QFI ratio $\mathcal{R}(\omega)$. Panels (a) and (b) show $\mathcal{F}(\omega)^\mathrm{Heterodyne}/Q(\omega)$ for the polynomial case with $s{=}3$ and $s{=}4$, respectively. As shown in the figure, the performance of heterodyne detection decreases as the nonlinearity $\beta$ increases. Specifically, at the point where the QFI ratio $\mathcal{R}(\omega)$ reaches its maximum---\textit{corresponding to the highest degree of \red{nonlinear-enhanced} frequency estimation attainable for the given probe with a specific} $s$ and time $\beta t$---heterodyne detection captures only a fraction of the available information. For $s = 3$, this fraction is approximately $0.3$, while for $s = 4$, it drops significantly to about $0.02$. Panels (c) and (d) show the fraction $\mathcal{F}(\omega)^\mathrm{Heterodyne}/Q(\omega)$ for the generalized squeezing case with $s{=}3$ and $s{=}4$, respectively. As seen from the figure, for $s = 3$, this fraction is approximately $0.3$ (comparable to the polynomial case), while for $s = 4$, it drops to about $0.1$. Note that for the generalized squeezing case, the QFI ratio is significantly larger than that of the polynomial case. Nonetheless, as evidenced by the heterodyne performance, it is possible to achieve improved sensing capabilities for the generalized squeezing case using a feasible heterodyne detection scheme.

\section{V. Alternative Two-Step Strategy Encoding}

\begin{figure}
    \centering
    \includegraphics[width=0.8\linewidth]{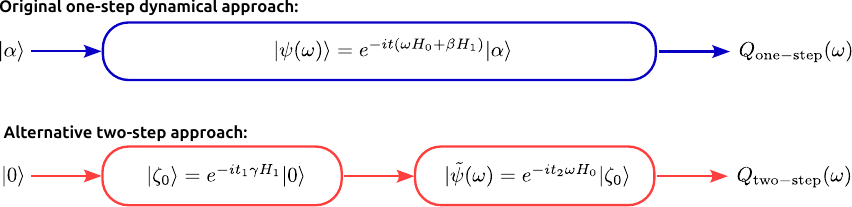}
    \caption{\vm{Two different strategies for encoding an unknown frequency $\omega$ into a quantum state for estimation. Top panel (blue): The original one-step approach proposed in this work, where a coherent state evolves under the full nonlinear probe Hamiltonian $H = \omega H_0 + \beta H_1$ for a total time $t$, dynamically encoding the unknown frequency. Bottom panel (red): An alternative two-step strategy in which the vacuum state is first evolved for a time $t_1$ under the nonlinear Hamiltonian $H_1$ alone, generating the intermediate state $|\zeta_0\rangle$. This prepared state then encodes the unknown frequency during a subsequent free evolution for time $t_2$ under $\omega H_0$. The two strategies are energy-matched by requiring both probes to have the same average energy. The goal is to compare their respective frequency estimation performance by evaluating the quantum Fisher information for each case, namely $Q_\mathrm{one{-}step}(\omega)$ for the one-step approach and $Q_\mathrm{two{-}step}(\omega)$ for the two-step strategy.}}
    \label{fig_schematic_alternative}
\end{figure}

\vm{An alternative two-step strategy for frequency sensing can be developed within our framework by decomposing the full dynamics into two distinct steps, see Fig.~\ref{fig_schematic_alternative} for a schematic. In the first stage, a state $|\zeta_0\rangle$ is prepared by evolving the field vacuum state $|0\rangle$ for a time $t_1$ using only the nonlinear part of the Hamiltonian, i.e.
\begin{equation}
|\zeta_0\rangle = e^{-i t_1 \gamma H_1} |0\rangle,
\end{equation}
where $H_1$ is one of the nonlinear Hamiltonian families shown in Eqs.~\eqref{eq_family_of_nonlinearities} and $\gamma$ is a nonlinear parameter. In the second stage, the unknown frequency $\omega$ is encoded by allowing the prepared state $|\zeta_0\rangle$ to evolve freely for a time $t_2$ under the action of $H_0=a^\dagger a$.}

\vm{To enable a fair comparison with the one-step sensing approach, we choose the parameter $\gamma$ such that the average energy of the alternative two-step probe equals that of the one-step coherent probe. Specifically, we determine the value $\gamma^*$ that ensures the intermediate state of the two-step scheme has the same energy as the initial state in the one-step setup, by imposing the condition:
\begin{eqnarray}
    \langle \zeta_0|\omega H_0|\zeta_0\rangle &=& \langle \alpha|\omega H_0 + \beta H_1|\alpha\rangle\,.
\end{eqnarray}
Hence, ensuring both strategies to have the same energy on average.}

\vm{Following the main results presented in the text, we now analyze two specific cases, namely the generalized squeezing case and the polynomial case for $s = 3$ and $s = 4$. In both scenarios, we fix the coherent amplitude to $\alpha = 1$ and set the times as $\beta(t_1 + t_2) = \beta t = 0.01$. We consider two different choices for the duration of the first step nonlinear evolution, that is $t_1 = 0.95t$ and $t_1 = 0.5t$. Once $t_1$ is fixed, the remaining time $t_2 = t - t_1$ for the free evolution is determined accordingly.}

\begin{figure}[t]
    \centering
    \includegraphics[width=0.9\linewidth]{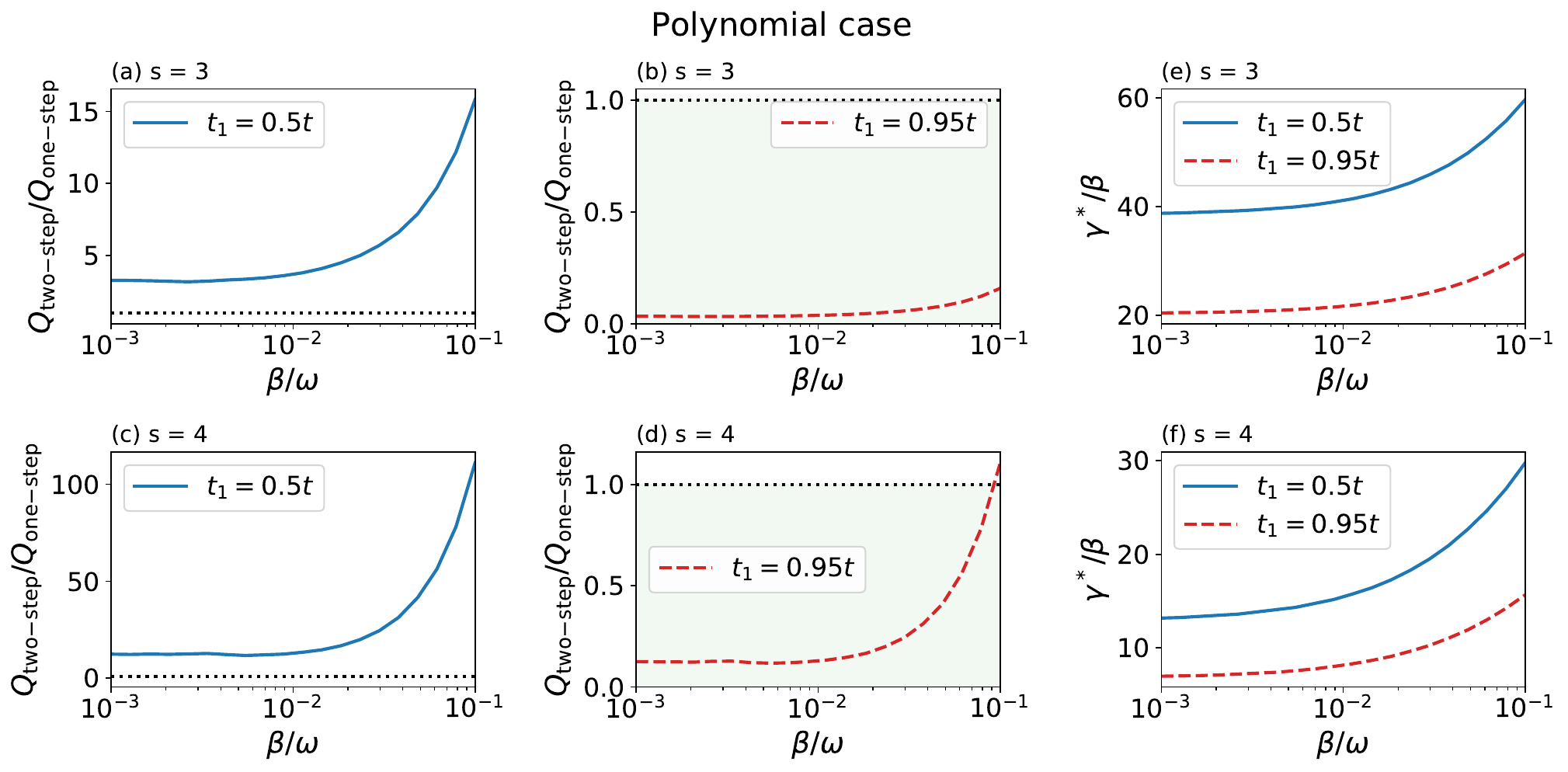}
    \caption{\vm{Comparison between the one-step and two-step strategies for the polynomial case with $s = 3$ and $s = 4$. Panels (a)–(d) show the ratio $Q_\mathrm{two{-}step}/Q_\mathrm{one{-}step}$ as a function of $\beta/\omega$ for two different choices of $t_1$. Panels (e)–(f) show the ratio $\gamma^*/\beta$ as a function of $\beta/\omega$ for $s = 3$ and $s = 4$. Other parameters are $\alpha = 1$ and $\beta t = 0.01$. Light green shaded region represents the region in which our strategy outperforms the two-step strategy.}}
    \label{fig_comparison_pol}
\end{figure}

\vm{In Fig.~\ref{fig_comparison_pol}, we compare the one-step and two-step strategies for the polynomial case with $s = 3$ and $s = 4$. Figs.~\ref{fig_comparison_pol}(a)–(d) show the ratio $Q_\mathrm{two{-}step}/Q_\mathrm{one{-}step}$ as a function of $\beta/\omega$ for two different choices of $t_1$. As these plots show, the one-step strategy significantly outperforms the two-step one (i.e., $Q_\mathrm{one{-}step} > Q_\mathrm{two{-}step}$) when more time is spent preparing the intermediate state $|\zeta_0\rangle$. Conversely, the two-step strategy becomes clearly superior (i.e., $Q_\mathrm{one{-}step} < Q_\mathrm{two{-}step}$) when less time is devoted to state preparation, allowing more time for free evolution and hence more information acquisition.}

\vm{Nonetheless, a key observation in this comparison is that the parameter $\gamma^*$, which is chosen to ensure both probes have the same average energy, is significantly larger than $\beta$. This is illustrated in Figs.~\ref{fig_comparison_pol}(e)–(f), where we plot the ratio $\gamma^*/\beta$ as a function of $\beta/\omega$ for $s = 3$ and $s = 4$. As the figures show, $\gamma^*$ consistently exceeds $\beta$ by a wide margin. This implies that, regardless of whether $Q_\mathrm{one{-}step} > Q_\mathrm{two{-}step}$ or $Q_\mathrm{one{-}step} < Q_\mathrm{two{-}step}$, the two-step strategy demands a much stronger nonlinearity $\gamma^*$, pushing it beyond the physically relevant regime considered in this work, where $\beta \leq \omega$. Consequently, our original one-step strategy is far more energy-efficient and experimentally feasible than the two-stage scheme.}

\begin{figure}[t]
    \centering
    \includegraphics[width=0.9\linewidth]{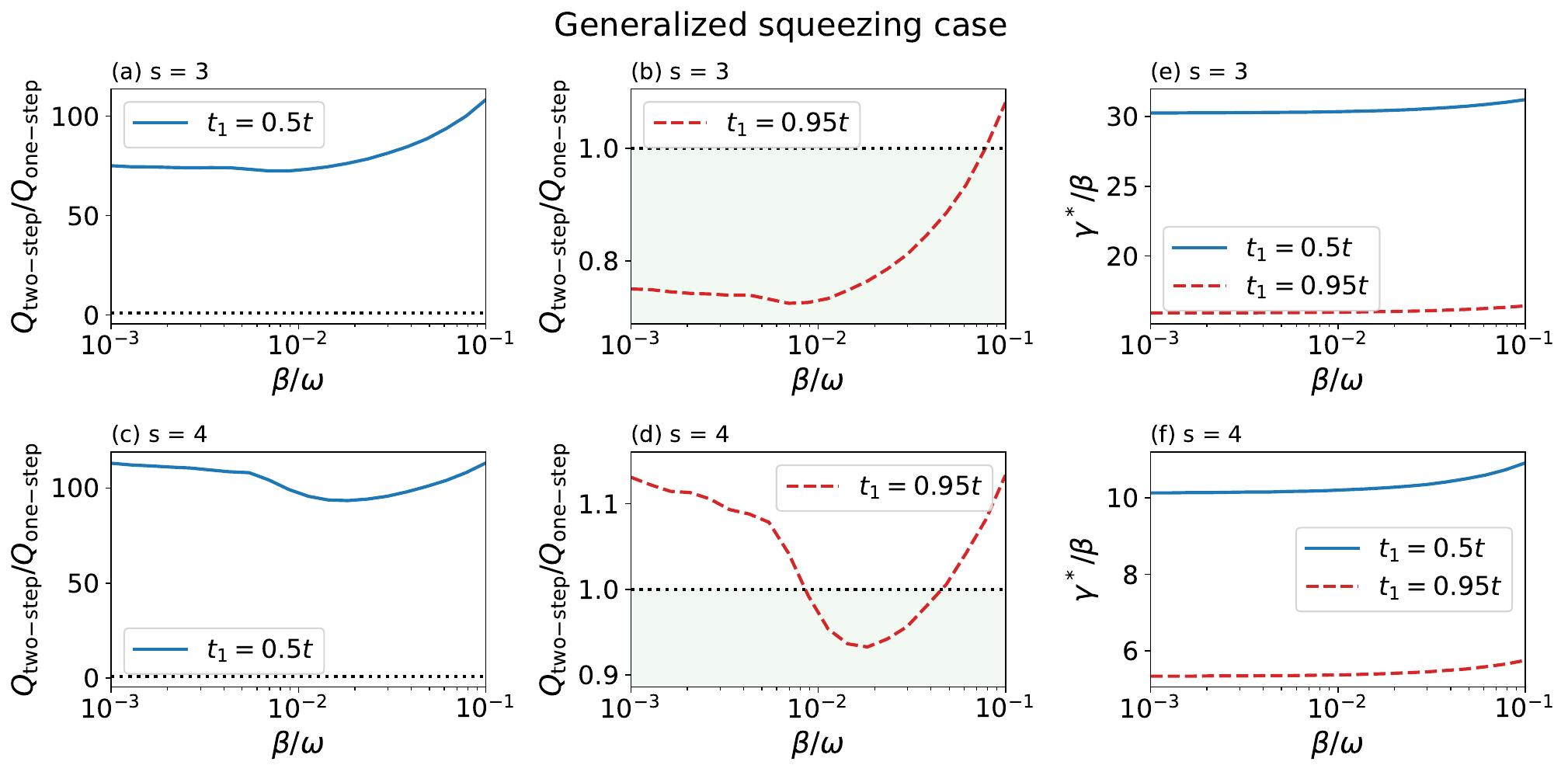}
    \caption{\vm{Comparison between the one-step and two-step strategies for the generalized squeezing case with $s = 3$ and $s = 4$. Panels (a)–(d) show the ratio $Q_\mathrm{two{-}step}/Q_\mathrm{one{-}step}$ as a function of $\beta/\omega$ for two different choices of $t_1$. Panels (e)–(f) show the ratio $\gamma^*/\beta$ as a function of $\beta/\omega$ for $s = 3$ and $s = 4$. Other parameters are $\alpha = 1$ and $\beta t = 0.01$. Light green shaded region represents the region in which our strategy outperforms the two-step strategy.}}
    \label{fig_comparison_sqz}
\end{figure}

\vm{Similarly, in Fig.~\ref{fig_comparison_sqz}, we compare the one-step and two-step strategies for the generalized squeezing case with $s = 3$ and $s = 4$. Figs.~\ref{fig_comparison_sqz}(a)–(d) show the ratio $Q_\mathrm{two{-}step}/Q_\mathrm{one{-}step}$ as a function of $\beta/\omega$ for two different choices of $t_1$. As shown in the figures, we observe the same trend as in the polynomial case, namely: the one-step strategy outperforms the two-step one (i.e., $Q_\mathrm{one{-}step} > Q_\mathrm{two{-}step}$) when more time is allocated to preparing the intermediate state $|\zeta_0\rangle$. In contrast, the two-step strategy becomes superior (i.e., $Q_\mathrm{one{-}step} < Q_\mathrm{two{-}step}$) when less time is spent on state preparation, leaving more time for free evolution.}

\vm{A similar observation holds here. The parameter $\gamma^*$, chosen to ensure both probes have the same average energy, is significantly larger than $\beta$. This is clearly illustrated in Figs.~\ref{fig_comparison_sqz}(e)–(f), where we plot the ratio $\gamma^*/\beta$ as a function of $\beta/\omega$ for $s = 3$ and $s = 4$. Across all cases, $\gamma^*$ consistently exceeds $\beta$ by a substantial margin. This reinforces the conclusion that our one-step strategy is not only more energy-efficient but also more experimentally feasible than the two-stage scheme.}

\vm{As a final remark, beyond the quantitative comparison, we now offer additional arguments in favor of our strategy. The one-step sensing approach is indeed more physically relevant, as it relies on a single-step, naturally \textit{stimulated} process rather than a two-step, carefully \textit{tailored} sequence. This simplicity not only makes our dynamical sensing approach more realistic in general but also enhances its practicality from an experimental implementation perspective.}

\vm{Consider the polynomial case with $s = 3$ as an example. Let $t_1$ denote the time spent preparing the initial non-Gaussian state, and let $t - t_1$ be the remaining time used for frequency encoding via free evolution. The state $U_3|0\rangle$ prepared during $t_1$ has an average photon number of $27(\gamma t_1)^2$. To ensure a fair comparison, this energy must match that of the one-step coherent probe, whose mean photon number is $\alpha_0^2$. Hence, we require $27(\gamma^* t_1)^2 = \alpha_0^2$. The quantum Fisher information for estimating $\omega$ using the two-step strategy is given by
\begin{equation}
Q_\mathrm{two{-}step}(\omega) = 4(t - t_1)^2 (\Delta a^\dagger a)^2,
\end{equation}
where $t$ is the total time used in our original one-step dynamical sensing approach. The quantity $(\Delta a^\dagger a)^2$ represents the fluctuations of the photon number operator, which must be evaluated with respect to the non-Gaussian state prepared during the time interval $t_1$, and explicitly can be casted as:
\begin{equation}
    (\Delta a^\dagger a)^2 = 9 (\gamma^* t_1)^2[7 + 864(\gamma^* t_1)^2].\label{fluctuations_magic}
\end{equation}
Note that Eq.~\eqref{fluctuations_magic} (and thus the quantum Fisher information) is independent of $\omega$, yet $\gamma^*$ depends on the choices of $\{\omega, \beta,\alpha\}$. In addition, one can readily evaluate the ratio $\mathfrak{R}_\mathrm{alt}(\omega)$ for the two-step sensing strategy as follows:
\begin{equation}
    \mathfrak{R}_\mathrm{alt}(\omega) = \left[\frac{7}{3}+288\left(\frac{\gamma^*}{\beta} [\beta t_1]\right)^2\right]\left[1 - \frac{t_1}{t}\right]^2.
\end{equation}
From the above, we can simply take $\beta t_1 = 0.5 \beta t$, corresponding to one of the time allocations considered in Figs.~\ref{fig_comparison_pol} and \ref{fig_comparison_sqz}. Substituting this into our expression for the quantum Fisher information ratio yields:
\begin{equation}
\mathfrak{R}_\mathrm{alt}(\omega) = \frac{1}{4} \left[\frac{7}{3}+72\left(\frac{\gamma^*}{\beta} [\beta t]\right)^2\right],
\end{equation}
for the specific case illustrated in the Figs.~\ref{fig_comparison_pol} and \ref{fig_comparison_sqz}, where $\beta t = 0.01$, one gets:
\begin{equation}
\mathfrak{R}_\mathrm{alt}(\omega) = \frac{1}{4} \left[\frac{7}{3}+72\left(\frac{\gamma^*}{\beta}\right)^2 10^{-4}\right].
\end{equation}
A similar conclusion holds for the case $s = 4$, where the non-Gaussian state $U_4|0\rangle$ has an average photon number of $240(\gamma^* t_1)^2$---$\gamma^*$ accounts for probes with same energy. Finally, we emphasize that regardless of whether the two-step strategy performs better or worse than our original one-step coherent probe, the two-step scheme requires a nonlinearity $\gamma^*$ that is far beyond what is experimentally feasible. In fact, the required values of $\gamma^*$ lie well outside our current parameter regime, where $\beta \lesssim \omega$.}
\end{document}